\documentclass[usenatbib]{mn2e}
\usepackage[dvips]{graphicx}
\usepackage{graphicx}
\usepackage{amsmath,amssymb}
\usepackage{epsf}
\usepackage{dcolumn}
\usepackage{color}
\usepackage{epsf}
\usepackage{psfrag}
\usepackage{epsfig}
\usepackage{bm}

\newcommand{\be}{\begin{equation}}
\newcommand{\ee}{\end{equation}}

\begin{document}
\title[Number Counts and Dynamical Vacuum Cosmologies]{Number Counts and Dynamical Vacuum Cosmologies}
\author[Devi, Borges, Carneiro \& Alcaniz]
{N. Chandrachani Devi$^1$\thanks{E-mail:chandrachani@on.br.},~
H. A. Borges$^{2,3}$\thanks{E-mail:humberto@ufba.br},~
S. Carneiro$^{2}$\thanks{E-mail: saulo.carneiro@pq.cnpq.br},~ 
and
J. S. Alcaniz$^{1}$\thanks{E-mail: alcaniz@on.br}
 \\
$^1$Observat\'orio Nacional, 20921-400, Rio de Janeiro, RJ, Brazil\\
$^2$Instituto de F\'{\i}sica, Universidade Federal da Bahia, 40210-340, Salvador, BA, Brazil\\
$^3$Institute of Cosmology and Gravitation, University of Portsmouth, PO1 3FX, Portsmouth,
UK\\
} 

\date{\today}

\maketitle

\begin{abstract}
We study non-linear structure formation in an interacting model of the dark sector of the Universe in which the dark energy density decays linearly with the Hubble parameter, $\rho_{\Lambda} \propto H$, leading to a constant-rate creation of cold dark matter. We derive all relevant expressions to calculate the mass function and the cluster number density using the Sheth-Torman formalism and show that the effect of the interaction process is to increase the number of bound structures of large masses ($M \gtrsim 10^{14} M_{\odot}h^{-1}$) when compared to the standard $\Lambda$CDM model. Since these models are not reducible to each other, this number counts signature can in principle be tested in future surveys. 
\end{abstract}


\section{Introduction}

With the accumulation of high-quality cosmological data from type Ia supernovae (SNe Ia) observations, anisotropies in the cosmic microwave background (CMB) and measurements of the large-scale properties of the Universe (LSS), the need for an acceleration mechanism governing the late-time cosmic evolution 
has been confirmed \citep{Sahni2000,Peebles2003,Padmanabhan2003, Copeland2006,Alcaniz2006,Frieman2008}. From the observational viewpoint, the standard $\Lambda$CDM model, whose cosmic dynamics is driven by a cosmological constant $\Lambda$ and a component of cold dark matter (CDM), remains as the most consistent cosmological setting. However, despite its observational successes, the standard model suffers from fundamental theoretical issues (see, e.g.\citep{Weinberg1986,Sahni2000}), which motivate the investigation of various alternative scenarios such as quintessence \citep{Caldwell1998,Ratra1988,Liddle1999,Steinhardt1999}, modified gravity theories \citep{Carroll2004, Capozziello2005,Amendola2007,Fay2007,Santos2007}, decaying vacuum models \citep{Ozer1986,Ozer1987, Bertolami1986,Freese1987,Carvalho1992,Shapiro2009,Alcaniz2005,Costa2010}, among others (see~\citep{Peebles2003, Padmanabhan2003, Copeland2006,Alcaniz2006,Frieman2008} and refereces therein).

The implementation of observational tests that are able to distinguish between such scenarios and the standard one has become one of the most important tasks nowadays in Cosmology since they not only test the observational viability of these alternative models but also discuss the standard cosmology in a more general perspective. However, many of these models behave very similarly to the standard $\Lambda$CDM paradigm at the level of background evolution, making it difficult to distinguish them through observational data such as SNe Ia measurements. This forces us to go beyond the background expansion, i.e., towards perturbations and the non-linear regime of structure formation. In this regard, observation of cluster number counts as function of the cluster mass and redshift have been accounted as a potential candidate for probing both the expansion rate and the growth of perturbations, thereby providing independent constraints on cosmological parameters, such as the present-day matter density 
parameter 
$\Omega_{m0}$ and the rms density fluctuation $\sigma_{8}$. 
This kind of data can, therefore, provide a new  window for cosmological modelling, with the ability of distinguishing between the various dark energy models or among the various alternative scenarios by their effects on structure formation. It is also worth mentioning that there are a number of surveys, either on going or being planned for the near future, that are expected to release a large number of cluster counts data. This certainly will improve the complementarity of cluster observations with other cosmological probes, such as CMB, SNe Ia and LSS observations.

In this paper we study the non-linear evolution of density perturbations in a spatially flat cosmological model in which vacuum decays linearly with the Hubble parameter, leading to the production of cold dark matter at a constant rate \citep{Borges2005}. This is a particular case of models with interaction in the dark sector \citep{Zimdahl2001,Costa2008, Wu2008,Jesus2008,Campo2009,Koyama2009,Chimento2010,He2011,Wangs2012} that has been shown competitive, with good concordance in both background and linear perturbation levels \citep{Carneiro2006,Carneiro2008,Borges2008,Pigozzo2011,Alcaniz2012, Velten2013}. Further, we analyze how such an interacting process affects the predicted halo abundance. For this purpose we calculate all relevant expressions for the mass function and the cluster number density following  the Sheth-Torman formalism \citep{Sheth1999} and compare our results with that of the standard $\Lambda$CDM scenario. We also test the robustness of our results by considering a mass function different from the one introduced by Sheth-Torman. The structure of the paper is as follows. In Section 2, we introduce the decaying 
vacuum model and discuss the background evolution. In Section 3, we study non-linear density perturbations through the spherical collapse model. We calculate the number 
density as function of 
the cluster mass using the Sheth-Torman formalism in Section 4 and finally  we discuss our main results in Section 5. We are using $8\pi G = c = 1$ in this paper.

\section{Background}

We consider a FLRW cosmology in which there is a continuous decay of vacuum energy into dark matter. In other words, the dark sector consists of cold dark matter interacting with a varying dark energy component with equation-of-state parameter $\omega = -1$. Owing to this interaction, total energy conservation is expressed by the continuity equation as
\begin{equation} 
\dot{\rho}_{m} + 3\frac{\dot{a}}{a}\rho_{m} = - \dot{\rho}_{\Lambda},
\label{cont}
\end{equation}
where $\rho_{m}$ and $\rho_{\Lambda}$ are the energy densities of CDM and vacuum, respectively, and the Friedmann equation reads
\begin{equation}
\label{eq.fried}
3 H^2(a)\equiv \Big(\frac{\dot{a}}{a}\Big)^2=\rho_{\rm{m}}+{\rm{\rho_{\Lambda}}}\;.
\end{equation} 
Some possible forms of the time-dependent vacuum have been investigated in the literature \citep{Ozer1986,Ozer1987, Bertolami1986,Freese1987,Carvalho1992,Alcaniz2005,Shapiro2009,Costa2010}. In this paper we consider a particular ansatz $\rho_{\Lambda} \propto H(a)$, which corresponds to a constant-rate CDM creation from vacuum \citep{Borges2005}. With this ansatz we derive the Hubble expansion as a function of the redshift as
\begin{equation} \label{hz}
H(z)=H_{0}[1-\Omega_{m0}+\Omega_{m0}(1+z)^{3/2}].
\end{equation}
Since matter is no longer conserved, it does not evolve as $\rho_{m}(z) = 3H_0^2 \Omega_{m0} (1+z)^3$ as in the standard model, but rather as 
\begin{equation}
\rho_{m}(z) = 3H_0^2 \Omega_{m0} \left[ \Omega_{m0} (1+z)^{3}+ (1-\Omega_{m0})(1+z)^{3/2} \right].
\label{eq:rhom}
\end{equation}
The second term on the right-hand side of Eq.~(\ref{eq:rhom})~results from dark matter creation. At high redshifts, the above scaling relation reduces to 
\begin{equation} \label{highz}
\rho_{m}(z) = 3H_0^2 \Omega_{m0}^2 (1+z)^3 \quad \quad \quad \quad (z \gg 1),
\end{equation}
where the extra factor $\Omega_{m0}$ means that, for the same amount of matter at high-$z$, the decaying model predicts a larger amount at present when compared to the $\Lambda$CDM model. {In fact, the test of this model against observations of the matter linear power spectrum  led to $\Omega_{m0} \approx 0.45$ (\citet{Borges2008, Alcaniz2012}). This value is in good agreement with results from background observations, namely the Hubble diagram for type Ia supernovae (SNIa), the position of the first acoustic peak in the CMB anisotropy spectrum, and baryonic acoustic oscillations (BAO) data \citep{Pigozzo2011}.}  For a more detailed discussion on the current  observational constraints on this class of models, we refer the reader to \citet{Alcaniz2012}.

\begin{figure*}
\vspace{0.5cm}
\psfrag{m}[c][c][1][0]{\Large {\bf ${\rm log_{_{10}}}(M/h^{-1}M_{\odot})$}}
\psfrag{n}[c][c][1][0]{\Large ${\rm log_{_{10}}}[dn/d{\rm log_{_{10}}}M/(h^{-1}{\rm Mpc})^{-3}]$}
\psfrag{dn}[c][c][1][0]{\Large $\frac{(dn/d{\rm log_{_{10}}}M)_{\Lambda(t)}}{(dn/d{\rm log_{_{10}}}M)_{\Lambda}}$}
\includegraphics[width=14cm,height=10cm]{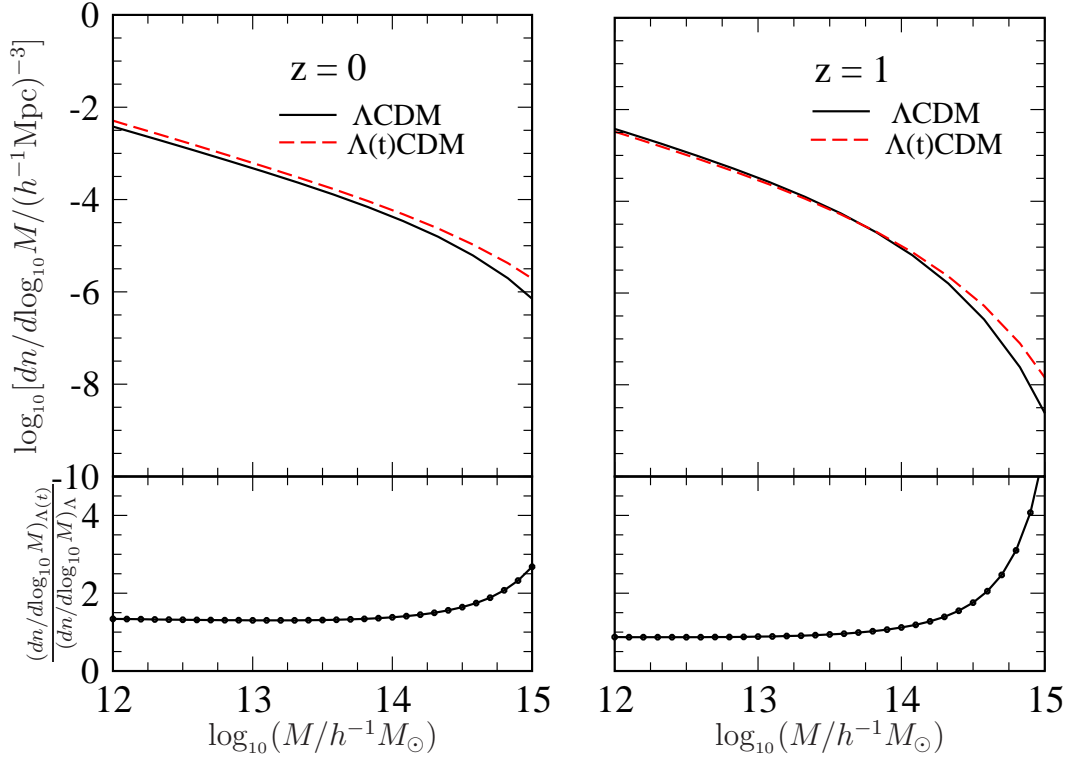}
\caption{{\it {Upper panels}}: The comoving number density as a function of mass for the decaying vacuum and the standard $\Lambda$CDM model at redshifts $z = 0$ and $z = 1$, respectively. {\it Lower panels:} Corresponding differences in the comoving number density with respect to the standard $\Lambda$CDM model.}
\label{fig:Nodensity}
\end{figure*}

\section{Evolution of matter perturbations}
In this section, we analyze the basic equations required to study the behavior of matter perturbations within the framework of the above discussed cosmological model.
Let us start with the equations for energy conservation, momentum conservation and the gravitational field,
\begin{equation}\label{o}
\frac{\partial\rho}{\partial t}+\nabla_r\cdot(\rho\vec u)+p\nabla_r\cdot\vec u=0,
\end{equation}
\begin{equation}\label{t}
\frac{\partial\vec u}{\partial t}+(\vec u\cdot\nabla_r)\vec u=-\nabla_r\phi-\frac{\nabla_r p}{\rho+p},
\end{equation}
\begin{equation}\label{tr}
\nabla_r^2\phi=\frac{\rho+3p}{2},
\end{equation}
where $\rho$, $p$, $\vec u$ and $\phi$ are, respectively, the energy density, pressure, velocity and gravitational potential of the cosmic fluid. Introducing the cosmological perturbations 
\begin{equation}\label{pert1}
\rho_m(\vec r,t)=\rho_m(t)+\delta\rho_m(\vec r,t),
\end{equation}
\begin{equation}\label{pert2}
\vec u(\vec r,t)=\vec u_0(t)+\vec v(\vec r,t),
\end{equation}
\begin{equation}\label{pert3}
\phi(\vec r,t)=\phi_0(t)+\Phi(\vec r,t)
\end{equation}
into the set of equations (\ref{o})-(\ref{tr}), we obtain
\begin{equation}\label{r1}
\dot{\delta\rho_m}+3H\delta\rho_m+\rho_m\nabla_r\cdot{\vec v}+(\nabla_r\cdot\vec v)\delta\rho_m+(H\vec r+\vec v)\cdot{\nabla_r\delta\rho_m}=0,
\end{equation}
\begin{equation}\label{r2}
\dot{\vec v}+H(\vec r\cdot{\nabla_r})\vec v+H\vec v+(\vec v\cdot{\nabla_r})\vec v+\nabla_r\Phi=0,
\end{equation}
\begin{equation}\label{r3}
\nabla_r^2\Phi=\frac{\delta\rho_m}{2},
\end{equation}
where $\vec u_0=H\vec r$ corresponds to the Hubble flow and $\vec v$ corresponds to the matter peculiar velocities. Here we have taken $\delta \rho_{\Lambda} = 0$, which is valid for the model discussed in Sec. 2 when one considers modes well inside the horizon \citep{Zimdahl2011}. Generally, in the spherical collapse model we assume a homogeneous distribution of matter within a sphere of radius $\vec r=a(t)\vec x$, so the term $\nabla_r\delta\rho_m$ appearing in the energy conservation equation is null. Furthermore, to maintain a spherically symmetric profile we choose a velocity field of type $\vec v = \theta \vec x/3$, with $\nabla . \vec v =\theta $. We then use the definition of the density contrast $\delta_m=\delta\rho_m/\rho_m$ and change to comoving variables, with $\nabla_r=\nabla/a$. Hence, equations (\ref{r1}) and (\ref{r2}) are rewritten as
\begin{equation}\label{4r}
\dot\delta_m=-\frac{\Psi}{\rho_m}\delta_m-\frac{\theta}{a}(1+\delta_m),
\end{equation} 
\begin{equation}\label{5r}
\dot\theta+H\theta+\frac{\theta^2}{3a}=-\frac{a\rho_m\delta_m}{2},
\end{equation} 
where $\Psi=-\dot{\rho}_{\Lambda}$.
Differentiating $(\ref{4r})$ with respect to cosmic time and eliminating $\theta$ and $\dot\theta$, we obtain a second-order non-linear differential equation for the matter density contrast,

\begin{align}
\label{res}
\ddot\delta_m &+\left(2H+\frac{\Psi}{\rho_m}\right)\dot\delta_m \nonumber \\ 
&+\left[\frac{d}{dt}\left(\frac{\Psi}{\rho_m}\right)+2H\frac{\Psi}{\rho_m}-\frac{\rho_m(1+\delta_m)}{2}\right]\delta_m \\
&=\frac{1}{3(1+\delta_m)}\left(4\dot\delta_m^2+\frac{5\Psi}{\rho_m}\delta_m\dot\delta_m+\frac{\Psi^2}{\rho_m^2}\delta_m^2\right).\nonumber 
\end{align}

The corresponding linear equation is obtained for $\delta_m\ll 1$ and is given by \citep{Arcuri1994,Borges2008}
\begin{equation}\label{reslinear}
\ddot\delta_m+\left(2H+\frac{\Psi}{\rho_m}\right)\dot\delta_m+\left[\frac{d}{dt}\left(\frac{\Psi}{\rho_m}\right)+2H\frac{\Psi}{\rho_m}-\frac{\rho_m}{2}\right]\delta_m=0. 
\end{equation}
Changing variables from $t$ to redshift $z$, the evolution equation (\ref{res}) takes the form
\begin{eqnarray}
\label{oi}
&&H^2(1+z)^2\delta_m''-\left[f(z)(1+z)^2+H(1+z)\frac{\Psi}{\rho_m}\right]\delta_m' \nonumber \\
&& + \left[2H\frac{\Psi}{\rho_m}-\frac{1}{2}\rho_m(1+\delta_m)\right]\delta_m = \frac{4}{3}H^2(1+z)^2\frac{\delta_m'^2}{1+\delta_m}\nonumber \\
&&-\frac{5}{3}H(1+z)\frac{\Psi}{\rho_m}\frac{\delta_m\delta_m'}{1+\delta_m}+\frac{1}{3}\left(\frac{\Psi}{\rho_m}\right)^2\frac{\delta_m^2}{1+\delta_m}, 
\end{eqnarray}
where the background functions are given by Eqs. (\ref{hz})-(\ref{eq:rhom}) and
\begin{subequations}
\begin{equation}\label{ps0}
\frac{\Psi}{\rho_m}=\frac{3H_0}{2}(1-\Omega_{m0}),
\end{equation}
\begin{eqnarray}
\label{pug}
f(z)&=&-\frac{H_0^2}{2}(1+z)^{-1}[2\Omega_{m0}+\Omega_{m0}(1+z)^{3/2}-2] \nonumber \\ 
&&\times \left[1-\Omega_{m0}+\Omega_{m0}(1+z)^{3/2}\right].
\end{eqnarray}
\end{subequations}

\section{Number counts}

In this section we study the imprint of the particle creation process described above on the predicted abundance of bound objects, such as clusters and groups of galaxies. The formation of these structures can be understood by the spherical top-hat model, in which density perturbations initially increase linearly with the scale factor in the matter dominated phase, according to Eq. (\ref{reslinear}). Once the over-density exceeds a critical value $\delta_c$, it decouples from the Hubble flow, reaches to a maximum radius and then starts collapsing under its own gravitational force until its virialization to a bound structure. The critical density contrast $\delta_c$ is a model-dependent quantity and we calculate it for the model of Sec. 2 as follows~\citep{Abramo2007,Pace2010,Campanelli2011}.  First we integrate the non-linear equation (\ref{res}), looking for the initial conditions for which $\delta_m$ diverges at some redshift $z_c$. Using these same initial conditions we integrate the linear equation (\ref{reslinear}) to obtain $\delta_c(z_c)$. We also obtain the linear growth function solving Eq.(\ref{reslinear}) with the initial conditions $\delta_i \sim a_i$ and $\delta_i^{\prime} \sim 1$ at some initial $a_i$ (say, at the epoch of matter-radiation equality).

The comoving number density of collapsed objects at redshift $z$ within the mass interval $M$ and $M+dM$ is given by
\begin{equation} 
\frac{dn(M,z)}{dM} = - \frac{\bar{\rho}_{m}}{M}\frac{d \ln \sigma(M,z)}{dM}f(\sigma),
\label{eq:dNdz}
\end{equation}
where $\bar{\rho}_{m}$ is the comoving background density,  $\sigma(z)$ represents the variance of the linear density field on a given comoving scale $R$ and $f(\sigma)$ is the mass function. 
Note that in the $\Lambda$CDM case $\bar{\rho}_{m}$ is equal to the present density $\rho_{m0}$, but, owing to (\ref{eq:rhom}), for the interacting model discussed here it is given by $\bar{\rho}_{m}=\rho_{m0}\beta(z)$,  with $\beta(z)= \Omega_{m0}+(1-\Omega_{m0})(1+z)^{-3/2}$.  In what concerns the mass function, it is well known that the Press-Schechter formalism~\citep{Press1974} presents over (under) predictions at the low (high) mass limit when compared with numerical simulations. Thus, other mass functions $f(\sigma)$ have been proposed in the literature (see e.g. \cite{Sheth1999,Jenkins2001,Reed2003,Reed2007,Tinker2008}). In our analysis we use a modified form of the Press-Schechter mass function, the so-called Sheth \& Torman mass function, which is motivated by the ellipsoidal collapse of overdense regions and is more compatible with present-day simulations. It reads as \citep{Sheth1999}
\begin{equation} \label{fsigma}
f(\sigma) = A\sqrt{\frac{2q}{\pi}}\left[1+\left(\frac{\sigma^2(z)}{q\delta_c^2(z)}\right)^p\right]\frac{\delta_c(z)}{\sigma(z)}\exp\left[- \frac{\delta_c^2(z)q}{2\sigma^2(z)}\right],
\end{equation}
where $A$ is a normalization constant. Values of ($p$, $q$) equal to ($0$, $1$) recover the Press-Schechter mass function, whose shape and amplitude are motivated by the spherical (rather than ellipsoidal) collapse of matter overdensities. The values $p=0.3$ and $q=0.707$ correspond to an accurate fitting of the $\Lambda$CDM model with N-body simulations \citep{Sheth1999}.

\begin{figure}
\vspace{0.5cm}
\psfrag{m}[c][c][1][0]{\large {\bf ${\rm log_{_{10}}}(M/h^{-1}M_{\odot})$}}
\psfrag{dn}[c][c][1][0]{\large ${\rm log_{_{10}}}[dn/d{\rm log_{_{10}}}M/(h^{-1}{\rm Mpc})^{-3}]$}
\includegraphics[width=80mm]{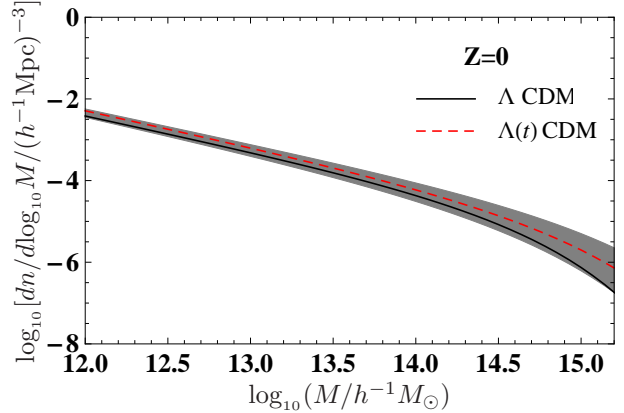}
\caption{Variation of the comoving number density with respect to mass for the decaying vacuum (dashed line) and the standard $\Lambda$CDM model (solid line). The shadowed region corresponds to the comoving number density of the decaying vacuum model when the Sheth-Torman parameters $(p, q)$ vary about $30\%$ from their standard values.}
\label{fig:stvary}
\end{figure}

The variance $\sigma(z)$ can be written as \citep{Bond1991,Eke1996,Peebles1993,Dodelson2003}:
\begin{equation}
\sigma^2(R,z) = \frac{D^2(z)}{2\pi^2}\int_0^{\infty} k^2 P(k)W^2(kR) dk,
\end{equation}
with a corresponding mass $M=\frac{4\pi}{3}\rho_{m0}R^3\beta(z)$ instead of the standard-case $M=\frac{4\pi}{3}\rho_{m0}R^3$. The growth function $D(z) =\delta(z)/\delta(0)$ is normalized such that $D(z)=1$ at present. $W(kR)$ is the Fourier transform of a spherical top-hat filter of radius $R$, i.e.
$
W(kR) = 3 \left(sin(kR)/{(kR)^3}-cos(kR)/{(kR)^2}\right).
$
The CDM power spectrum is given by $P(k)=P_{0} k^{n_s} T^2(k)$, where $T(k)$ is the transfer function. We use the BBKS transfer function \citep{Bardeen1986,Martin2000} which, for a baryon density parameter $\Omega_{b0}\ll\Omega_{m0}$, can be approximated by
\begin{eqnarray} \label{transfer}
&&T(x=k/k_{eq})=\frac{ln[1+0.171x]}{(0.171x)}\times \\ \nonumber
&&\left[1 + 0.284x + (1.18x)^2 + (0.399x)^3 + (0.490x)^4\right]^{-0.25},
\end{eqnarray}
where $k_{eq}^{-1}$ is the comoving Hubble scale at the redshift of matter-radiation equality. Note that, because of the matter density scaling (\ref{highz}), for the decaying vacuum model we obtain $k_{eq}=0.073h^2 \Omega^2_{m0} M{\rm pc}^{-1}$ instead of the standard result $k_{eq} = 0.073h^2\Omega_{m0}M{\rm pc}^{-1}$ (we set the present-day radiation density parameter as $\Omega_{R0} = 4.15 \times 10^{-5} h^{-2}$). The normalization of the power spectrum is done in terms of $\sigma_8$, the present rms fluctuation at a scale of $8h^{-1}$Mpc. With this, the expression for $\sigma^2(M,z)$ becomes  
\begin{equation} \label{sigma}
\sigma^2(M,z) =\sigma_{8}^2\frac{D^2(z)}{D_{\Lambda}^2(0)}\frac{\int_0^{\infty} k^{n_s+2} T^2(\Omega_{m0},k)W^2(kR)dk}{\int_0^{\infty} k^{n_s+2} T_{\Lambda}^2(\Omega_{m0\Lambda},k)W^2(kR_8)dk}\;,
\end{equation}  
where the index $\Lambda$ stands for the $\Lambda$CDM model. We set $\sigma_{8} =0.83$, $h = 0.67$ and $n_s= 0.971$, as given by the~\citet{2013arXiv1303.5076P}. It is worth mentioning that, for our interacting scenario, the concordance value of the matter density parameter obtained from a joint analysis involving CMB, SNe Ia, BAO and the linear power spectrum is about $\Omega_{m0} = 0.45$~\citep{Carneiro2006,Carneiro2008,Borges2008,Pigozzo2011,Alcaniz2012}. We will use this value in the further analyses of this model, whereas for the $\Lambda$CDM case we fix the matter density parameter at $\Omega_{m0\Lambda} = 0.3$.

\section{Discussion}

In Figs. 1a and 1b (upper panels) we show the comoving number density as function of the mass $M$ for $z=0$ and $z = 1$, respectively. In both cases, the interacting model predicts approximately the same  number density as the standard model for cluster masses below $M = 10^{14} M_{\odot}$.  The corresponding differences in the number density with respect to the $\Lambda$CDM case are depicted in the {lower panels} of Fig 1. We find that the predicted $dn/dM$ difference between these scenarios increases considerably for $M > 10^{14} M_{\odot}$, reaching $\sim 38\%$ and $\sim 64\%$, respectively, for $M = 10^{14} M_{\odot}h^{-1}$ and  $M = 10^{14.5} M_{\odot}h^{-1}$ at $z = 0$ and $\sim 12\%$ and $\sim 74\%$ at $z = 1$. This is an important result since ongoing and planned surveys, like eROSITA~\citep{erosita2012,Pillepich2012} and J-PAS~\citep{Benitez2014}, must be able to detect such differences and, therefore, to distinguish between this class of interacting cosmology and the $\Lambda$CDM model. 

As mentioned earlier, the predicted behavior of $dn/dM$ depends  on the values of the parameters $p$ and $q$ [see Eq. (\ref{fsigma})]. In order to better  visualize this  dependence, we show in Fig. 2 the expected change in the evolution of the
comoving number density  considering different values of the pair ($p$, $q$) within a 30\% interval (gray region) from the standard values $p = 0.707$ and $q = 0.3$ discussed in the previous section (red dashed line). We find that larger values of $p$ and $q$ moves the interacting model prediction towards the standard result. For completeness, we also tested the dependence of these results with the Sheth-Torman mass function by repeating our calculations using the mass function introduced in ~\citet{Reed2003,Reed2007}. No significant difference from the results shown in Fig. 1 was found. From these results, we conclude that the main difference in cluster number density between the  $\Lambda$CDM and interacting models is due to the dynamical behavior of the models instead of the mass function used in the number counts calculations. {The reader may also ask whether the small difference between the predictions of the two models is just due to the different best-fit values of the matter density parameter, $\Omega_{m0} = 0.3$ for the standard model and $\Omega_{m0} = 0.45$ for the interacting model. An inspection of equation (\ref{sigma}), however, shows that this is not the case. For the $\Lambda$CDM model, a change in $\Omega_{m0}$ from $0.3$ to $0.45$ leads indeed to a small difference in the number density whereas in the interacting case, the matter density appears squared in the transfer function (\ref{transfer}), leading to a larger difference when it is changed from $0.45$ to $0.3$. This is illustrated in Fig. 3.}

\begin{figure}
\vspace{0.5cm}
\psfrag{m}[c][c][1][0]{\large {\bf ${\rm log_{_{10}}}(M/h^{-1}M_{\odot})$}}
\psfrag{n}[c][c][1][0]{\large ${\rm log_{_{10}}}[dn/d{\rm log_{_{10}}}M/(h^{-1}{\rm Mpc})^{-3}]$}
\includegraphics[width=80mm]{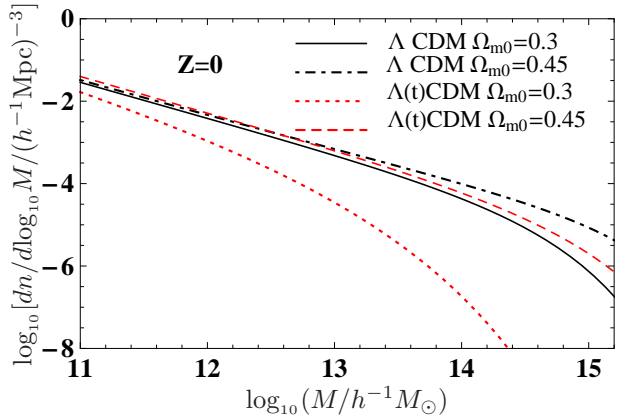}
\caption{Comoving number density for the standard $\Lambda$CDM and the interacting models with different values of the matter density parameters.}
\label{fig3}
\end{figure}

\section{Final Remarks}

The nature of the cosmological dark sector is certainly one of the main open problems of modern cosmology. A residual $\Lambda$ term, although in good agreement with current observations, exacerbates the well known cosmological constant problems~\citep{Weinberg1986}, requiring a natural explanation for its small value. In this paper we have discussed the ability of cluster number counts to constrain a class of alternative scenarios in which the vacuum energy density decays linearly with the Hubble parameter $H$. This particular class of $\Lambda$(t)CDM models has been  tested against precise background observations, namely the position of the first peak of the CMB anisotropy spectrum, the Hubble diagram for type Ia supernovas and the BAO distance scale \citep{Carneiro2006,Carneiro2008, Pigozzo2011} and also provides a good fit to the linear power spectrum of matter \citep{Borges2008}. It is worth mentioning that a joint analysis of these observations has led to a remarkable concordance, with $\Omega_{m0} \approx 0.45$ \citep{Alcaniz2012}, which is the matter 
density value used to perform the present results.

We have studied the structure formation in these $\Lambda$(t)CDM scenarios and derived all relevant expressions to calculate the mass function and the cluster number density using the Sheth-Torman formalism. We have found that the cluster counts prediction of the $\Lambda$CDM and $\Lambda$(t)CDM models are approximately the same for cluster masses below $M = 10^{14}$ solar masses and becomes different for larger values of $M$ and higher $z$. Such differences, however, are not large enough to be confronted to current observations but may be detected by ongoing and planned surveys capable to observe high mass galaxy clusters at higher redshifts (e.g., eROSITA~\citep{erosita2012, Pillepich2012} and J-PAS~\citep{Benitez2014}). We have also studied the dependence of our results on the Sheth-Torman mass function by performing a similar analysis with the mass function introduced in \citet{Reed2003,Reed2007}. We have found no significant difference in the results, which makes them robust. A quantitative comparison of these models predictions with 
the current cluster number observations and mock data from ongoing surveys is the subject of a forthcoming publication.

\section{Acknowledgement}
The authors thank T. Roy Choudhury for helpful discussions. This work is partially supported by CNPq, FAPERJ, FAPESB and INEspa\c{c}o.

\end{document}